# The promoters of human cell cycle genes integrate signals from two tumor suppressive pathways during cellular transformation

Running title: Human cell cycle gene promoter regulation

Yuval Tabach[1,2,5], Michael Milyavsky[1,5], Igor Shats[1,5], Ran Brosh[1,5], Or Zuk[2], Assif Yitzhaky[2], Roberto Mantovani[3], Eytan Domany[2], Varda Rotter[1] Yitzhak Pilpel[4]

1. Department of Molecular Cell Biology, Weizmann Institute of Science, Rehovot, 76100 Israel
2. Department of Physics of Complex Systems, Weizmann Institute of Science, Rehovot, 76100 Israel
3. Dipartimento di Scienze Biomolecolare e Biotecnologie, Universita di Milano, Milan, Italy
4. Department of Molecular Genetics, Weizmann Institute of Science, Rehovot, 76100 Israel
5. These authors contributed equally to this work

Correspondence to:
Yitzhak Pilpel[4] Department of Molecular Genetics, Weizmann Institute of Science, Rehovot 76100, Israel. Tel.: +972 8 934 6058; Fax: +972 8 934 4108; E-mail: pilpel@weizmann.ac.il
Varda Rotter[1] Department of Molecular Cell Biology, Weizmann Institute of Science, Rehovot 76100, Israel. Tel.: +972 8 934 4501; Fax: +972 8 946 5265; E-mail: Varda.Rotter@weizmann.ac.il

Character Count = 75,000

**Abstract**

Deciphering regulatory events that drive malignant transformation represents a major challenge for systems biology. Here we analyzed genome-wide transcription profiling of an *in-vitro* transformation process. We focused on a cluster of genes whose expression levels increased as a function of p53 and p16$^{INK4A}$ tumor suppressors inactivation. This cluster predominantly consists of cell cycle genes and constitutes a signature of a diversity of cancers. By linking expression profiles of the genes in the cluster with the dynamic behavior of p53 and p16$^{INK4A}$, we identified a promoter architecture that integrates signals from the two tumor suppressive channels and that maps their activity onto distinct levels of expression of the cell cycle genes, which in turn, correspond to different cellular proliferation rates. Taking components of the mitotic spindle as an example, we experimentally verified our predictions that p53-mediated transcriptional repression of several of these novel targets is dependent on the activities of p21, NFY and E2F. Our study demonstrates how a well-controlled transformation process allows linking between gene expression, promoter architecture and activity of upstream signaling molecules.

**Introduction**

Cellular process are controlled by highly intricate regulatory networks (Bar-Joseph et al., 2003; Ihmels et al., 2002; Lee et al., 2002; Pilpel et al., 2001; Segal et al., 2003; Sharan et al., 2004; Sharan et al., 2003; Shen-Orr et al., 2002; Tavazoie et al., 1999; Werner, 2001). Most successes to date in understanding such networks were obtained in lower organisms; extension to mammalian genomes is complicated in part due to the complexity of the promoter and enhancer regions and also due to the tremendous intricacy of some of the regulatory circuits. Nevertheless, initial studies, e.g. in fly and in mammalian organisms succeeded in delineating promoter elements controlling specific to particular networks of genes (Berman et al., 2002; Berman et al., 2004; Elkon et al., 2003; Halfon et al., 2002; Smith et al., 2005; Sumazin et al., 2005; Thompson et al., 2004; Wasserman et al., 2000; Werner et al., 2003; Zhu et al., 2005). One such relevant study is that of Elkon et al. (Elkon et al., 2003) who pursued genome-wide *in-silico* identification of transcriptional regulators controlling the human cell cycle. Recent studies (Segal et al., 2003) explored an additional level in the signaling network in yeast, namely links between gene expression profiles and activity of signaling molecules that may directly affect transcription. Here too, extension to higher organisms is complicated by the considerable increase in the intricacy of signaling network architecture.

In addition to deciphering normal physiological processes, elucidation of regulatory and signaling networks is expected to allow better understanding of pathological conditions, such as cancer (Segal et al., 2004). Monitoring gene expression changes on a genome-wide scale is a powerful method to study transcriptional programs involved in carcinogenesis (Cho et al., 2001; Liotta and Petricoin, 2000; Whitfield et al., 2002). Indeed, specific expression signatures that correlate with specific diagnosis, survival, and response to therapy were proposed (Liotta and Petricoin, 2000; Rosenwald et al., 2003; Scherf et al., 2000). Yet, associations of those signatures with specific biological processes or with distinct genetic alterations acquired by cancer cells along *in-vivo* transformations are not obvious. The difficulties largely stem from different genetic backgrounds of patients, variable and uncharacterized mutations in tumors, and the uncontrolled contribution of inflammatory, endothelial, and stroma cells, which contaminate tumor specimens, to the measured gene expression patterns.

Thus, in order to obtain both novel and more reliable insights into genetic networks associated with oncogenesis, we have recently developed an *in-vitro* model for cellular transformation through a stepwise process (Milyavsky et al. 2003). The 600-day long transformation process (Fig. 1A) started with normal human diploid fibroblasts that entered replicative senescence after 40 population doublings (PDLs). In order to overcome replicative senescence, the cells were infected with human telomerase (hTERT), resulting in immortalization. After 150 PDLs following hTERT introduction, clones showed an increased proliferation rate. At that stage, cells lost expression of the $p16^{INK4A}$ and $p14^{ARF}$ tumor suppressors (INK4A locus) (Milyavsky et al., 2003). To explore the transcriptional and phenotypic impact of p53 at different stages of the transformation process, the p53 protein was inactivated by expression of a dominant-negative p53 peptide (GSE56) (Ossovskaya et al., 1996). The combination of these genetic manipulations in conjunction with H-ras over-expression gave rise to cells capable of forming tumors in nude mice (Milyavsky et al., 2005). Recent studies have described similar inactivation of $p16^{INK4A}$ in additional human cell lines that overcame telomere-independent crises during immortalization (Taylor et al., 2004; Tsutsui et al., 2002). Furthermore, using various experimental models it was shown that full transformation could be achieved by the combination of viral oncogenes together with cellular genes (Hahn et al., 1999; Voorhoeve and Agami, 2003). Collectively, these experimental models generated a model of defined genetic aberrations that initiate and promote the neoplastic process.

We have previously suggested that our *in-vitro* cellular system reproduces some of the distinct stages that characterize solid tumor initiation and progression (Milyavsky et al., 2005). In the present study we aimed at deciphering the specific transcriptional networks associated with defined stages of malignant transformation. Thus, we utilized genome-wide mRNA expression profiling recently carried out at 12 time points along the transformation process using the GeneChip Human Genome Focus Array (Affymetrix, Santa Clara, CA) that represents over 8,500 verified human sequences from the NCBI RefSeq database (Milyavsky et al., 2005). Subsequent cluster analysis of mRNA expression profiles identified ten stable clusters. One of them, termed the "proliferation cluster" (Milyavsky et al., 2005), showed a pronounced sensitivity to the status of p53 and $p16^{INKA}$ tumor suppressors. Importantly, a large number of genes found in this proliferation cluster also clustered in studies that analyzed human primary tumor samples (Alizadeh et al., 2000; Barrett et al., 2003; Perou et al., 2000; Rosty et al., 2005). In addition, the proliferation cluster genes were found to be, on average, significantly more highly expressed in tumors obtained from patients with bad outcome compared to patients with good outcome in breast cancer (Dai et al., 2005; Milyavsky et al., 2005) as well as in other cancers (Tabach et al., in preparation). These findings strongly support the notion that the proliferation cluster genes are highly relevant to naturally occurring cancers.

Here, we further analyzed the proliferation cluster in an attempt to reveal how the promoters of its genes generate a transcriptional program that integrates the activity of tumor suppressors within the cell. By linking expression profiles of the genes in the cluster with the dynamic behavior of p53 and $p16^{INK4A}$, we identified two promoter architectures that integrate different signals from the two tumor suppressive channels and that map their activity onto distinct levels of expression of the cell cycle genes, which in turn, correspond to different cellular proliferation rates.

**Results**

**The "proliferation cluster"**

Our recent cluster analysis of mRNA expression profiles during the cellular transformation process identified ten stable clusters (Milyavsky et al., 2005). According to the clustering method used (Blatt et al., 1996) a stable cluster is one which is robust against perturbing the data; on the one hand, the points that belong to it are (relatively) remote from other points, while, on the other hand, they constitute a well defined entity, i.e. a (relatively) contiguous region of high density of data points. The Super Paramagnetic Clustering algorithm we used is capable of grouping together points that constitute such a high density region, irrespective of the region's shape, and also provides a quantitative measure of the robustness or stability of our identification of these points as a "cluster".

Here we focus on one of these clusters, which was a-posteriori termed the "proliferation cluster" (due to its genes annotations, see below). The cluster has a somewhat elongated shape, yet it is a stable and a well defined cluster that cannot be naturally divided into sub-clusters (Fig. S1). We decided to focus on this cluster since the genes that constitute it showed a complex and interesting behavior, i.e. pronounced, yet non-trivial (see below), sensitivity to the status of both p53 and p16$^{INK4A}$ tumor suppressors (Milyavsky et al., 2005).

Fig. 1A shows a schematic description of the *in-vitro* transformation process. Fig. 1B shows expression profiles of the 168 genes of the proliferation cluster. Cells doubling rates for selected stages along the process are presented in the table below the expression matrix. Telomerase and p53 activity status are also shown. Four major states are distinguished. 1. The "young-cell period" (1$^{st}$ and 2$^{nd}$ columns in the expression matrix) corresponds to the early passages of cells and includes the first cell cycle divisions. During this period the cells are young and the expression profiles of the genes in the cluster are relatively high. At the second point during the young cell period (day 30) a separate cell culture was derived from the above culture by introduction of the p53 dominant-negative peptide, GSE56. The cluster genes responded by up-regulation (mean expression was significantly elevated from 406.5 to 479.5, P value = $3.1*10^{-23}$ by paired t-test). 2. The senescence period (3$^{rd}$ and 4$^{th}$ columns in the expression matrix) is characterized by arrest of cell divisions. The expression level of the genes in the cluster was dramatically decreased during this period. Yet, even within the senescence period, expression profiles of the genes in the cluster were significantly elevated at the 4$^{th}$ time point compared to the 3$^{rd}$ point, i.e. in response to p53 inhibition (p-value = $7.1*10^{-28}$ by paired t-test). 3. The "slow immortalization" period (5$^{th}$ - 7$^{th}$ columns in the expression matrix) is characterized by expression levels similar to that in young cells. 4. The "fast immortalization" stage (8th through 12th columns in the expression matrix), which was shown to be associated with silencing of p16 (Milyavsky et al., 2005), is characterized by a significant shortening of the cell cycle period (see table below expression matrix, Fig. 1B,). In parallel, a further increase in the expression levels of the cluster genes was observed. At the last two points of this period GSE56 was reintroduced, resulting in inhibition of p53. The genes in the cluster responded to p53 inhibition by significant (p-value = $1.4*10^{-32}$) up-regulation (column 11 and 12 in the matrix). It thus appears that the expression profiles of the genes in the proliferation cluster correlate with p53 activity as well as with the rate of cell proliferation.

We next examined functional annotations of the genes in the cluster. Out of 168 genes of the cluster, 112 have functional annotation in "Gene Ontology" (Ashburner et al., 2000). All enriched functional terms among these genes, along with statistical significance analysis (Dennis et al., 2003) are listed in Table S1. Notably, only cell cycle-related functions are significantly (p-value $<10^{-10}$) over represented in the cluster. A detailed examination of this gene cluster revealed that it includes mainly genes associated with various aspects of cell proliferation. We thus termed the cluster "the proliferation cluster". The genes in the cluster take part in diverse processes crucial for the transition through the different cell cycle phases, such as DNA replication (MCM2, MCM3, MCM5, MCM6, RRM1-2, RFC3-5, GMNN, POLA, POLD1, POLE, POLQ, PRIM1) and DNA repair (BLM,BRCA1,MSH6). G2/M phase genes represented the largest functional category. More specifically, cyclin-dependent kinase CDC2, whose function is critical for mitotic entry, and its regulators such as cyclin B2, CDC25A, and CDC25C, showed marked up-regulation. In addition, genes with distinctive function in mitosis, including mitotic spindle organization (CENPA, CENPF, TTK, BIRC5, kinesins), mitotic spindle checkpoint (BUB1, BUB1B, MAD2L1, CDC20), and chromosome segregation (PTTG1, CENPF, ESPL1, UBE2C, PLK1, STK12) were also up-regulated upon p53 inactivation. This cluster also includes genes that are responsible for DNA packaging (HAT1, CHC1, SUV39H1, TOP2A) and chromosome organization (H1FX).

Since many of the genes in the proliferation cluster belong to the core of the cell cycle machinery, we examined whether they display cell cycle periodicity using data from whole-genome mRNA profiles during HeLa cell divisions (Whitfield et al., 2002). We found that 53% of the genes in the proliferation cluster, for which expression data exists also in the HeLa cell cycle experiment, display a cell cycle periodicity (CCP) index (Tavazoie et al., 1999) greater than three (a score that corresponds to the top 6.7th percentile of all genes represented on the array). The expression profiles of these genes during cell cycle peak at the two main cell cycle checkpoints, namely the entry into the S and M phases (Fig. S2).

In summary, we found that the proliferation cluster is enriched with cell cycle periodic genes, whose functions are at the core of the cell cycle machinery. Expression of these genes peaks at the end of the transformation process in our *in-vitro* model, the only stage at which transplanted cells give rise to malignant tumors in mice. In addition, these genes are co-expressed in tumors derived from patients with various types of cancers, and show high correlation with poor outcome and prognosis (Milyavsky et al., 2005), attesting to their relevance to naturally-occurring cancers.

**Transcriptional regulation of the proliferation cluster genes**

We next turned to identify promoter regulatory motifs that drive the expression of genes in the proliferation cluster. Rather than attempting to discover *de-novo* promoter motifs, we assumed that transcription factors with previously elucidated binding sites may be involved in regulating the genes in the cluster. Therefore we searched within the promoters of the proliferation cluster genes for the presence of each of the 326 known vertebrate transcription factor binding sites, represented as position specific scoring matrices (PSSMs) in the MatInspector database (Quandt, 1995). For this analysis, we adopted a working definition of the promoter region as the genome sequence that spans 1000 base-pairs upstream of the transcription start site (TSS) of each gene (Materials & Methods). We used a gene-to-PSSM assignment algorithm as in (Elkon et al., 2003) in order to scan each PSSM against the promoters of all genes represented on the array. For each PSSM, we calculated a hyper-geometric p-value score (Hughes et al., 2000) to assess the extent to which a motif is over represented among the cluster's genes compared to the rest of the genes on the array. We used the Bonferroni correction criterion for multiple hypotheses testing to set a threshold for over-representation of motifs in the cluster. Noticeably, all significant motifs, apart from VMYB, and including NFY, E2F, CHR (Cell cycle genes Homology Region), ELK1 and CDE (Cell cycle-Dependent Element), are known to be involved in the regulation of cell cycle (Badie et al., 2000; Bracken et al., 2004; Buchwalter et al., 2004; Manni et al., 2001; Mantovani, 1998; Matuoka and Chen, 2002; Nevins, 2001). We therefore focused on these cell cycle motifs in all subsequent analyses. Table I. shows all PSSMs with statistically significant over-representation in the cluster (Table S5 shows statistics of all the motifs in MatInspector). We have also examined the presence of the motifs in the 5' UTRs of the cluster's genes and found only barely significant over-representations in the cluster (see Table S2), and have thus decided to concentrate on the upstream regions only in all further analyses.

**Evolutionary conservation of the motifs**

We examined promoters of mouse genes orthologous to the proliferation cluster genes and found that the same motifs are also significantly over-represented in these promoters compared to the promoters in the rest of the mouse genome (Table S3). We have further assessed conservation at an organization level beyond the mere presence/absence of motif, namely conservation of the motif architecture between the two species. We found considerable conservation at this level too, using two criteria: First, the combinations of motifs that regulate orthologous promoters were significantly more similar to each other compared to combinations of non-orthologs (Fig. S3, A). Second, we found a significant tendency to preserve the locations of the motifs relative to the transcription start site in orthologous promoters (Fig. S3, B-F). The high level of conservation observed attests to the functional role of the motifs in these promoters.

**Revealing a hierarchy of regulatory motif combinations**

Having found a set of regulatory motifs that likely control the expression of the genes in the proliferation cluster we next attempted to identify particular combinations of such motifs. Here and in subsequent analysis a motif was considered present in a promoter if it appeared in the motif's preferred distance relative to the transcription start site, as shown in Fig. 2 (and see Materials and Methods). To this end, we calculated the "synergy" (Banerjee and Zhang, 2003; Garten et al., 2005; Pilpel et al., 2001) and rate of co-occurrence (Elkon et al., 2003; Sudarsanam et al., 2002) between regulatory motifs. A pair of motifs was considered 'synergistic' if the extent of expression coherence (Lapidot and Pilpel, 2003; Pilpel et al., 2001) of genes containing both motifs in their promoters is significantly greater than that of genes containing either of the motifs alone (Pilpel et al., 2001). A pair of motifs was considered highly co-occurring if there is a significant overlap between the set of genes containing the two motifs, given the number of genes containing each motif, and the genome size (using a hyper-geometric test to assess significance of overlap). In a recent study in yeast, we showed that although the two measures need not necessarily correlate, they are largely congruent in reality (Garten et al., 2005). We found that NFY, E2F, CDE, and CHR show multiple mutual interactions with each other, many of which were supported by both synergy and co-occurrence. On the other hand, the ELK1 motif co-occurs significantly with E2F and CDE, yet it has no synergistic effect with other motifs on gene expression (Fig. 3A). Many of the interactions are observed also when only genes in the proliferation cluster are considered (Fig. 3A).

In order to gain more insight on such motif interactions we used the Combinogram analysis, which we have previously utilized to dissect regulatory networks in yeast (Pilpel et al., 2001). We searched for the above five regulatory motifs within the promoters of all varying genes represented on the array. Given the five regulatory motifs, we partitioned the genes represented on the array, for which we also had promoter sequence assignment, into up to $2^5 = 32$ non-overlapping gene sets, each defined by a unique binary signature that reflects the presence or absence of each of the five motifs in their promoters. In the Combinogram we group together genes with identical binary signatures, depict their motif content (in the binary black & white matrix), the similarity between the average expression profiles of all pairs of gene sets (upper part dendrogram), and the averaged expression profiles of the genes in each set (expression matrix at the bottom part). The Combinogram in Fig. 3B, which was obtained with 18 gene sets that were populated with genes (14 other potential sets were not populated with genes), reveals several clear trends. First, although the analysis was applied to the entire set of genes represented on the array, i.e. without any preceding clustering stage, a division (corresponding to the main branching point of the dendrogram, marked "1" in the dendrogram) into genes that contain some of the motifs, and genes that do not, appears. Gene sets that are to the left of branch point "1" largely represent the proliferation cluster signature, whereas gene sets that are on the right branch display more or less flat expression profiles with no particular trend. The motifs that appear to determine this split are mainly ELK1, or NFY & CHR. Genes that contain none of the five motifs (column #14 from left), display the flat profile, as do genes that contain one or two of the motifs, but not ELK1 (columns 11 through 18 from left). The left branch, that largely shows the proliferation signature, may be further divided (branch point "2") into gene sets that contain at least ELK1 (columns 1 through 8) vs. gene sets that contain NFY and CHR, but not ELK1. These differences in motif composition reflect themselves at the expression patterns – without ELK1 a clear decrease in expression in the senescence (40) and senescence GSE (4) time points ($3^{rd}$ and $4^{th}$ rows of the expression matrix) is seen, whereas an increase in expression in the last two time points is evident too. The presence of ELK1 appears to be both necessary and sufficient for its typical dictated expression pattern, genes that contain only ELK1 (and none of the rest of the four motifs, column 5) are members of the ELK1 cluster, and genes that do not contain ELK1 (columns 9-18) are not in the cluster. Branch point "3" further sharpens the ELK1-dictated signal, as ELK1 show an interesting interaction with NFY. Genes that contain ELK1 and NFY (columns 6 through 8) are located to the right of branch point "3", as they display an intermediate between the pure ELK1 pattern and the NFY&CHR pattern, while genes to the left of this branch point display a distinct pattern. In general this analysis shows a sticking correspondence between motif content and gross and fine differences in expression, akin to a previous, yet much simpler observation, made in yeast (Pilpel et al., 2001). It allows the dissection of the role of individual motifs and their combinations (e.g. the effect of CHR on the background of NFY and CDE is clear from the difference in expression patterns between columns 10 and 11 in which CHR is present and absent respectively). This analysis strongly suggests that indeed the five regulatory motifs examined here indeed govern gene expression profiles during the 600-day long malignant transformation process and that the proliferation signature represents a genuine response dictated by these motifs.

Next, we examined whether it is possible to trace the regulatory effect of these motifs down to the relatively microscopic time-scale of single cell cycle divisions. We therefore tested whether the motif architecture we discovered here also governs the expression of these genes during cell cycle divisions. To this end, we constructed a Combinogram based on the five motifs together with cell cycle expression data derived from the HeLa cell cycle experiment described above ((Whitfield et al., 2002), Fig. 3C). Notably, we observed similar relationships between motif combinations and their effects on expression in both the transformation and the cell cycle experiments. As in the transformation process, in the cell cycle too, NFY appears to interact synergistically with CHR, an interaction that gives rise to a clear G2-M phase expression pattern. Here the presence of the CDE motif further amplifies this pattern. The resemblance between the transformation and the cell cycle experiments indicates that the transcriptional regulation of the cluster during the complex and largely uncharacterized 600-day long transformation process can be reduced to the more 'atomistic' level of the regulation of cell cycle. Interestingly though, in the cell cycle experiment we did not detect any clear pattern dictated by E2F, alone, or through combinations with other motifs, suggesting either that despite intensive research on this transcription factor, an accurate description of its binding site is still missing, or that its regulatory role is too complex and diverse (Bracken et al., 2004). This too is consistent with a recent observation (Elkon et al., 2003) that although E2F is over-represented in the promoters of cell cycle genes, it is not restricted to genes that peak at specific cell cycle phase. Likewise, the ELK1 motif does not seem to affect gene expression throughout the cell cycle. The observation that the E2F and ELK1 motifs affect transcription mainly in the transformation experiment and less so in cell cycle may indicate that their role in the transformation process is not mediated through a direct effect on the cell cycle, but rather on a potentially higher level. Another potential explanation for the fact that presence of either ELK1 motif or the combination NFY and E2F was significant in the transformation experiment but not in HeLa cell cycle experiment may be related to the fact that in HeLa cells, both p53 and pRb are inactivated by the viral oncoproteins E6 and E7. Thus, if the ELK1 motif or the combination of NFY and E2F potentially mediate growth restrictive effects of these major tumor suppressive pathways, we would not expect these effects to be manifested in HeLa cells.

**The proliferation cluster genes integrate information from two tumor suppressive channels**

Our knowledge of the detailed molecular history of the transformation process in the *in vitro* model allowed us to extend our analysis beyond the formation of links between regulatory motifs and expression profiles. Since the activity of upstream tumor suppressors was manipulated and monitored during the transformation, we could link gene expression patterns, mediated by various regulatory motifs, to expression levels of these signals. The activity levels of two prime tumor suppressor genes,

p53 and p16, varied throughout the transformation process. Since p53 was inactivated at the protein level, we used as a surrogate for its activity, array-derived mRNA measurement of its regulated target, p21, which is indeed down regulated after each application of the p53 dominant negative peptide, GSE56.

First, we observed that while the averaged mRNA profiles of the genes in the proliferation cluster do not correlate with the mRNA levels of either p21 or p16 alone, they show high negative correlation (r = -0.85, the probability of pairs of random genes that are summed up as with p21 and p16 to obtain such correlation or lower with the proliferation cluster's average, is <1% as estimated by 10,000 random samples of pairs) with a profile obtained by summing the mRNA expression profiles of these two genes (Fig. 4A). Simple logical functions such as AND or OR gates were unable to describe accurately the thresholded activity levels of the genes in the cluster as a function of the mRNA concentration of p16 and p21 (using various thresholds on their activity). Instead, the relationship can be accurately modeled with a "sum-gate", namely the summated profiles of p21 and p16. The proliferation cluster genes thus appear to sum up the expression levels of the two tumor suppressors and produce an analog output in the form of expression profiles. Recently, similar sum-gated promoter architectures were observed in E. coli (Kalir and Alon, 2004; Setty et al., 2003).

More importantly, we then identified the promoter motifs that likely mediate such integrative function. Based on our results with the proliferation cluster, we analyzed separately all the genes represented on the array that contain in their promoters the ELK1 motif, and all genes that contain a combination of NFY and CHR (The two motif combinations that dictate distinct expression patterns throughout he transformation process). Fig. 4B shows that the CHR&NFY-regulated genes clearly depend on both tumor suppressors, and their expression levels map the presence/absence of the two suppressors onto four distinct expression levels (multiple t-tests on all six pairwise comparisons always yielded significant p-value, with the least significant p-value equal to $0.0014*10^{-4}$). In contrast, the ELK1 motif mainly mediates a response to the presence/absence of p16. The expression of ELK1-regulated genes is significantly up-regulated following p16 inactivation. Although the ELK1 transcription factor was previously implicated in the regulation of expression of proliferation genes (Gille et al., 1995; Ullrich and Schlessinger, 1990), its potential regulatory interaction with either p16 or p53 was not addressed before.

**Three-way linkage of expression, promoter architecture and tumor suppressor activity**

In order to gain further insights into the relationship between mRNA expression profiles and promoter architecture, we sorted the proliferation cluster genes using SPIN (Tsafrir et al., 2005), a sorting algorithm that positions genes with similar expression profiles in adjacent rows of an expression matrix (Fig. 5A). We also examined the presence of the regulatory motifs in promoters of genes in the cluster along the sorted expression matrix (Fig. 5C and D). Interestingly, the CDE and E2F motifs are relatively evenly scattered along the cluster. Taken together with the observation that they are significantly highly specific to the proliferation cluster suggest that these motifs are major constituents of the entire cluster. On the other hand, the CHR motif, and to a smaller extent the NFY motif, are mainly concentrated in promoters of genes in the "upper" part of the sorted cluster, while ELK1 mainly shows a preference towards the genes in the "lower" part. We thus conclude that while CDE and E2F constitute the cluster and are present in the majority of its genes, CHR, NFY and ELK1 serve to modulate the general expression patterns dictates by CDE and E2F.

Although the averaged expression profile of the cluster genes is strongly negatively correlated with the summated expression profiles of p16 and p21, but not with the individual tumor suppressors (Fig 4A), it is still possible that part of the genes only correlate with p21 while others only correlate with p16. We have thus measured (Fig. 5B) for each gene in the (sorted) cluster the correlation of its mRNA expression profile during the transformation process with the expression profiles of p16, p21 and with the summated profiles of p21 p16 (Fig. 5B). Strikingly, for the majority of the genes in the cluster the negative correlation with the summated profile is stronger than with the individual tumor suppressors. This is predominantly true for the genes in the "upper" part of the sorted cluster. Although for these genes there exists also a negative correlation with p21 alone, they are more strongly (negatively) correlated with the sum-gate, suggesting that the motifs regulating these genes are indeed integrating, by summing up, the information from the two suppressor channels. The correlation with p21 gradually diminishes with genes that are located towards the "lower" part of the cluster and on the other hand the correlations with p16 level show the opposite trend, namely a high correlation with genes in the lower part of the cluster. We stress that the data suggest no obvious point where the cluster can be sub-divided in two clusters based on correlations with the two tumor suppressors.

It was found recently that DNA binding activity of NFY transcription factor is positively regulated by CDK2 phosphorylation. This may explain the higher sensitivity of NFY-containing genes to p21 level as it specifically inhibits CDK2 (Hahn and Weinberg, 2002; Sherr, 1996; Sherr and Roberts, 1999; Weinberg, 1995). On the other hand, p16 specifically inhibits CDK4 and CDK6 (Hahn and Weinberg, 2002; Sherr, 1996; Sherr and Roberts, 1999; Weinberg, 1995). Thus, the increased sensitivity of ELK1-containing promoters to p16 levels enables us to propose novel role for CDK4/6 in ELK1 regulation.

The integration of these findings together with published experimental data allowed us to propose a network linking three layers of data – mRNA expression, promoter regulatory motifs/transcription factors, and the upstream tumor suppressors and signaling molecules (Fig. 5E). It is entirely possible though that additional tumor suppressors and transcription factors participate in the network and future analysis may reveal their identity and role.

**Experimental validation of computational predictions**

Our data suggested that the proliferation cluster genes are potential targets of p53- and p16-mediated transcriptional repression. Notably, many cluster genes including TOP2A, CCNB2, CCNA2, BIRC5, CDC2, CDC25C, PRC1, POLD1, PLK and others were previously shown to be downregulated by p53 (Burns et al., 2003; Hoffman et al., 2001; Krause et al., 2000; Li et al., 2004; Manni et al., 2001; St Clair et al., 2004; Tang et al., 2001; Wang et al., 1997; Yamamoto et al., 1994; Yun et al., 1999), validating our analysis and enabling us to propose numerous novel p53-transrepression targets. Interestingly, multiple components of the kinetochore complex and most of the known spindle checkpoint genes are found in our proliferation cluster. Since p53 was not previously implicated in the regulation of this group of genes, we decided to test for p53-mediated transcriptional repression of several genes from this functional category. Importantly, the regulatory network we proposed, based on the microarray experiment conducted under basal unstressed conditions, is expected to hold for cases where the upstream tumor suppressors are induced either by forced overexpression or by stress. We therefore tested whether a stress-induced p53 will repress the expression of several kinetochore/spindle genes. To this end, we treated normal and GSE56-expressing WI-38 cells with doxorubicin, a DNA-damaging agent and a potent p53 activator, and measured the levels of several proliferation cluster-derived genes by quantitative real-time PCR (qPCR). Confirming our hypothesis, we found that following DNA damage, Cdc20, Bub1, CCNF, and Mad2L1, all of which are cluster members, were downregulated in normal WI-38 cells, but not in their isogenic counterparts, in which p53 was inactivated (Fig. 6). Thus, these kinetochore- and spindle checkpoint- related genes represent novel targets of p53-mediated transcriptional repression.

Since our computational analysis revealed that the proliferation cluster genes display a negative correlation with p21 mRNA profile, we tested whether p53 exerts repression of proliferation cluster genes via p21 induction. To this end, we treated the HCT-116 colon carcinoma cells and their p53-null and p21-null derivatives with doxorubicin. We measured the expression levels of several proliferation cluster genes by qPCR and calculated the fold repression for each gene as the ratio of expression level in non-treated cells to that in doxorubicin-treated cells (Table II). Notably, only cells that contained both functional p53 and p21 (HCT-116 p53+/+) displayed down-regulation of these genes following DNA-damage. This supports the notion that the proliferation cluster genes are transcriptionally repressed by p53, and suggests that this repression is mediated through p21.

In order to gain further insights into the mechanism of p53-dependent repression of the proliferation cluster genes, we decided to focus our efforts on the cdc20 gene as a representative member of the cluster. We cloned the cdc20 promoter into a luciferase reporter vector and transfected it into HCT-116 p53-/- cells with or without a p53 expression plasmid. As indicated in Fig. 7A, in the presence of wild-type p53, the activity of cdc20 promoter was significantly repressed. In contrast, a p53 mutant lacking a functional transactivation domain (L22Q/W23S) did not repress cdc20. The requirement for a functional transactivation domain supports our conclusion that cdc20 repression by p53 is indirect and is mediated by induction of a mediator gene. Co-transfection of cdc20 promoter reporter with the p16 expression vector also resulted in repression of promoter activity (Fig. 7A), supporting our prediction that the proliferation cluster genes integrate signals from both p53 and p16.

Since promoters of the proliferation cluster genes are highly enriched with E2F motifs, we tested whether cdc20 promoter activity is affected by the presence of an E2F1 dominant negative protein (E2F-dTA) that is capable of DNA binding but defective in its transactivation and RB-binding domain. Overexpression of this construct displaces the endogenous E2F proteins from the DNA, abolishing both activation and repression by E2F family members (Hofmann et al., 1996). As demonstrated in Fig. 7B, cdc20 promoter activity decreased in the presence of a dominant-negative E2F1, and p53 did not repress it further under those conditions (see Figure legend for details). The most significantly enriched motif in the proliferation cluster promoters is NF-Y, suggesting the involvement of its cognate transcription factor in the regulation of the cluster's genes. To validate this hypothesis, we tested whether cdc20 reporter activity is affected by the presence of an NF-Y dominant negative protein (Mantovani et al., 1994). Indeed, overexpression of a dominant negative NF-YA (dnNF-YA) resulted in reduction of cdc20 promoter activity and in strong attenuation of the p53-dependent repression of this promoter. These results indicate that both the E2F family of transcription factors and the NF-Y transcription factor participate in cdc20 regulation, and that p53-dependent repression of cdc20 is mediated through these proteins.

Finally, we addressed the significance of the NF-Y motifs found in the cdc20 promoter for p53-mediated repression. Two NF-Y motifs reside in cdc20 promoter within the first 100 bp relative to the TSS. We generated cdc20 promoter reporter constructs that harbor mutations in each of the motifs and an additional construct with both NF-Y motifs mutated. These constructs, together with the wild-type promoter, were tested for their responsiveness to p53 status by cotransfecting them into HCT-116 p53+/+ cells in the presence or absence of a dominant-negative p53. While mutation of each NF-Y site alone did not affect p53-mediated repression (data not shown), mutations in both NF-Y motifs resulted in significant attenuation of the repression (Fig. 7C). These results support our computational prediction that NF-Y motifs, enriched in the promoters of the cluster genes, are involved in the regulation of these genes by p53.

**Discussion**

This study describes the analysis of genome-wide expression profiles of an *in-vitro* transformation process. Focusing on a well-defined expression cluster that consists predominantly of core cell cycle genes, we identified promoter motifs and their combinations that regulate the transformation process. We suggest that at least part of such regulation can be explained by a direct effect on cell cycle progression. Working with a controlled transformation process allowed us, for the first time, to not only establish a connection between gene expression and promoter architecture, but also to identify links to the activity of upstream tumor suppressors. Such a three-way connection was most revealing as it identified promoter motifs that likely "count" the number of active tumor suppressive channels and map them onto distinct expression states. Finally, detailed experimental analyses of selected genes experimentally established many of the suggested components of the network.

It is well known that activation of p53 leads to induction of p21 and inhibition of CDK2 activity (Hahn and Weinberg, 2002; Sherr, 1996; Sherr and Roberts, 1999; Weinberg, 1995). As depicted in Fig 5E, CDK2 controls E2F indirectly (through inactivation of RB by phosphorylation) and NFY directly (through CDK2-mediated phosphorylation). The CHR (cell cycle genes homology region) and the CDE (cell cycle-dependent element) represent "orphan" binding sites as the factors that bind these motifs are still un-cloned (Zwicker et al., 1995). These two elements are often found in close proximity to each other and these CDE/CHR "tandems" were shown to be important for the cell cycle-dependent expression of many genes. However, not always these two sites appear together. For example, a single CHR was shown to control cell cycle-dependent transcription of the cdc25C phosphatase gene and to cooperate with E2F or Sp1/3 sites (Haugwitz et al., 2002). Our genome-wide analysis strongly suggests the existence of a novel strong functional cooperation between CHR and NFY elements. According to our Combinogram analysis, the presence of these two sites in the proximal 200 bp upstream of the transcription start site is sufficient to dictate a G2/M expression pattern of multiple genes (Fig. 3C). In addition, we discover here that the same promoter architecture, namely the combination of NFY and CHR, is responsible for the integration of inputs from p21 and p16 during the *in-vitro* transformation process.

ELK1 transcription factor is a well-known downstream target of the MAP kinase pathway. It was demonstrated that proliferative inputs from deregulated MAP kinase pathway are counteracted by a negative feedback loop involving p16 activation with subsequent inhibition of CDK4/6 activities (Lin et al., 1998; Serrano. M et al., 1997; Zhu et al., 1998). Interestingly, our results indicate a strong negative correlation between the activities of ELK1-containing promoters and the expression level of p16, suggesting a possibility that p16 inhibits the activity of ELK1. To the best of our knowledge, this relationship was not reported previously. Since p16 specifically inhibits CDK4 and CDK6, it is possible that phosphorylation by these kinases plays a role in ELK1 regulation.

The two tumor suppressors studied here, namely p53 and p16, mainly respond to intrinsic and environmental signals, respectively. Thus, the promoter architecture discovered in this study integrates internal and external signals that affect core cell cycle genes. Such integration is performed by identifying and counting the number of active suppressive channels and mapping the result onto distinct expression levels. The intermediate expression level states, which correspond to precisely one active suppressive channel, may represent an "undecided" state. Such a state might be followed by either the high or low expression states of the cell cycle genes that may ensue after inactivation or activation of the second channel, respectively. Residing in such intermediate state can facilitate more rapid transition to one of the two extreme stages in response to addition or removal of intrinsic or environmental suppressive signal. In this respect it is crucial to note that the expression levels of the cluster's genes are correlated with proliferation rate (Fig. 1B); thus, promoter tuning of transcription of at least some the genes may affect proliferation. Many genes in the proliferation cluster represent previously identified targets of p53-mediated transcriptional repression. Our results significantly broaden the list of potential p53 transrepression targets. Here, for example, we identified an entire set of kinetochore/spindle genes, the expression of which is negatively regulated by p53. The functional significance of this finding is still unclear but it is tempting to speculate that loss of this transcriptional control contributes to aneuploidy formation, which is frequently found in tumors with mutated p53.

An additional important conclusion of our study relates to the mechanism of p53-mediated transcriptional repression. Unlike transactivation by p53, which clearly requires p53 binding to the regulatory sequences of targets, the mechanisms of repression by p53 are less well understood. The promoters of repressed genes usually do not contain the p53 consensus binding sites. Various mechanisms of p53 transrepression were proposed for different genes. These include interference with the functions of activators either involving p53 binding to DNA or through protein-protein interactions, direct interference with the basal transcription machinery, or recruitment of histone deacetylases and chromatin remodeling (for review see (Ho and Benchimol, 2003)). In addition, it was recently demonstrated for several genes that p53-mediated transcriptional repression requires the induction of p21 (Lohr et al., 2003).

Our study is the first to address the question of the mechanism that underlines p53-mediated repression in a more systematic way, using a three-way linkage of gene expression, promoter architecture and tumor suppressor activity. Our results strongly suggest that transcriptional repression by p53 is in most cases indirect and is mediated by p21 induction, which is transduced to E2F/CDE, NF-Y, and CHR motifs in the promoters of target genes. In addition we propose that overexpression of "proliferation signatures", found in a variety of aggressive human cancers, is frequently a consequence of p16 and p53 tumor suppressor inactivation.

Finally, it is very crucial to note that the proliferation signature described here has clear relationship with naturally occurring human tumors. Rosty and colleagues (Rosty et al 2005) have identified a cluster of genes (also named "proliferation cluster"), whose expression levels were predictive of outcome in samples derived from human patients with cervical cancer; low levels of expression characterized a subset of the patients with good outcome. In our previous work (Milyavsky et al., 2005), we have shown that there is a significant overlap between our proliferation cluster and that reported by Rosty et al., and we mentioned there other publications that reportded that many of our proliferation cluster genes constitute very good predictors of relapse vs. favorable outcome. These results demonstrate that the present proliferation signature, albeit obtained in vitro, has a strong predictive value regarding aggressive tumor behavior. However, we are aware of the fact that such common features should be carefully evaluated using additional types of naturally occurring malignancies. In addition, in the future, similar transformation processes, performed with additional cell lines and settings may be important for further establishing the generality of the signatures derived here. In that respect we note that in our previous work (Milyavsky et al., 2005) we addressed this issue by monitoring similar molecular events, such as INK4A locus inactivation in two additional cultures, thus supporting the generality of our findings.

**Materials and Methods**

**Promoter sequence**

DNA sequences upstream of human ORFs were downloaded from the GoldenPath server at UCSC http://genome.ucsc.edu/goldenPath/hg16/bigZips/. Putative regulatory regions (1000 bps upstream to transcription start site) for the different genes were obtained. We used for the original experiment (Milyavsky et al., 2005) the GeneChip Human Genome Focus Array (Affymetrix, Santa Clara, CA) that represents over 8,500 verified human sequences from the NCBI RefSeq database. We identified promoters for 8110 genes out of the 8,500. Of the 8110 genes we have selected 5582 genes that had a "present call" (according to Affymetrix calling procedure) in the two duplicates of at least one sample. Of the 168 genes in the proliferation cluster, 141 probe sets had a promoter in GoldenPath.

When more than one probe set on the array corresponded to same genomic locus (e.g. due to alternative splicing) we considered the corresponding regulatory region only once.

While the present analysis covers only the 8,500 genes represented on the GeneChip Focus Array that was used in our original experiment (Milyavsky et al., 2005), we have also examined the promoter motif content of all ~33,000 genes that were represented on the U133 Array (Affymetrix, Santa Clara, CA). We found additional 2316 genes that were not represented on the Focus Array that contain at least 2 of the discovered transcriptional modules; 36 of them contain 4 of the motifs analyzed here, see Table S4. These genes may represent additional candidates for the network discovered here.

**A collection of Position Specific Scoring Matrices (PSSMs)**

We used the MatInspector library of 326 PSSMs maintained by Genomatix (Release 4.1) (Quandt, 1995) and a customary promoter to PSSM assignment score (Elkon et al., 2003). We have then identified a threshold on this score, above which a PSSM is considered assigned to a promoter. For that we used the genes in a cluster and for a range of potential values of the threshold score we calculated, using the hyper-geometric statistic, the groups specificity score (Hughes et al., 2000) of the motif relative to the genes in the cluster. We identified and adopted the threshold score that minimizes the hyper-geometric probability function. See Fig. S5 for examples for threshold score determination for a selection of motifs. Only motifs that passed the Bonferroni correction for multiple hypotheses testing (that considered the multiple attempted thresholds) were retained.

**Assessing motif positional bias**

Positional bias was previously defined as the extent to which a motif that is assigned to a set of promoters is enriched in a sequence window (defined in terms of distance relative to the TSS) of a fixed length (e.g. 50 bps) with the maximal number of promoter (Hughes et al., 2000).Although efficient and simple, this algorithm has a limitation of having to define a fixed length window, without a priori knowledge about the relevant window width. We thus devised the following alternative procedure that learns the window's width from the data. We search for the window that is most enriched with occurrences of the motif using the following procedure:

1. Assume we have *N* occurrences of the motif in the promoters of the cluster's genes. Denote their (ordered) distances from the TSS (of each gene) by $C_i$ (i=1..N), such that : $0 \leq C_1 \leq .. \leq C_N < 1000$.

2. A window is defined as a subinterval $[a,b] \subset [0,1000]$. We search for the window most enriched with motifs, compared to a random background model. For each window [a,b], we denote by M(a,b) the number of motifs $C_i$ with distance of at least a and no more than b from the TSS.

The background distribution of the number of motifs in the window [a,b] is M(a,b) ~ Binomial(N, (b-a+1)/1000). Since windows overlap, an enrichment of a specific window leads to enrichment of windows overlapping it. Thus, we defined the most enriched window to be the one with smallest background probability, that is the interval [a,b] minimizing Pr(M(a,b)) under this background model. Obviously, the densest window is $[C_i, C_j]$ for some i≤j, therefore we can restrict our search only for intervals of the form $[C_i, C_j]$. Thus, it is defined as $W_{min}$ = $argmin_{i,j} Pr(M(C_i,C_j) \geq j-i+1)$, with $P_{min} = min_{i,j} Pr(M(C_i,C_j) \geq j-i+1))$.

3. To test statistical significance of the densest window, the distribution of $P_{min}$ in the background model is required. This was calculated by simulations. For N from 2 to 300 we have performed 100,000 simulations, each time selecting *N* points randomly in [0,1000] and then computing $P_{min}$. This gave an empirical distribution denoted $F_{N,min}$. The p-value for the observed most enriched window is simply $F_{N,min}(P_{min})$.

**Combinogram analyses**

The analysis was initiated with a set of N motifs. Each of the 5582 genes was assigned with a binary signature of length N with a 1 at the i$^{th}$ position if the gene contains motif i in its promoter, and a 0 otherwise. Thus, $2^N$ gene sets (that constitute the 'power set' of the set N), termed Genes defined by Motif Combinations (GMCs) were generated where all the genes in a given GMC share the same subset of the N motifs. The averaged expression profile of all the genes in each GMC was determined. The normalized Euclidian distance between averaged expression profiles for all pairs of GMCs was calculated and served as the input for the dendrogram analyses that were generated with the Cluster Analysis module in Matlab 7 (Mathworks) using the average-linkage option.

**Cell Lines**

Primary human embryonic lung fibroblasts (WI-38) were maintained in MEM supplemented with 10% fetal calf serum, 1mM sodium-Pyruvate, 2mM L-Glutamine, and antibiotics. Cells were passaged and counted once in five to seven days. Population doublings (PDLs) were calculated using the formula: PDLs = log(cell output/cell input)/log2. Doubling rates for the indicated culture stages were calculated by dividing the PDLs by the corresponding number of days in culture. The HCT-116 colon carcinoma cell line and its p53-null and p21-null derivatives were a gift from B.Vogelstein (The John Hopkins University, Baltimore, MD) and were described in (Bunz et al., 1998). HCT-116 cells were maintained in McCoy's medium supplemented with 10% fetal calf serum, 1mM sodium-Pyruvate, 2mM L-Glutamine, and antibiotics. All cell lines were grown at 37°C in a humidified atmosphere of 5% $CO_2$ in air.

**Plasmids**

The luciferase reporter construct, *p-cdc20-luc*, was generated by cloning cdc20 promoter and 5'-UTR into a luciferase reporter plasmid. Briefly, a genomic fragment of the cdc20 promoter, spanning from (-1002) bp to (+229) bp relative to the TSS, was amplified by PCR with the primers 5'-tccacctctgagcacattcat-3', 5'- tccttgcagttggtgcct-3', using Expand Long Template PCR system (Roche Applied Science). The amplified region was cloned into pGEM-T easy vector (Promega) and then transferred into pGL3 super basic vector (gift from M. Oren, Weizmann Institute of Science) using the restriction enzymes NdeI and NcoI. Mutations in NF-Y motifs were generated on the template of *p-cdc20-luc* using QuikChange Site–Directed Mutagenesis kit (Stratagene) together with the following primers (only sense primers are shown, mutations are in uppercase): for mutation of NF-Y motif at position (-83), CccttcgccggagaggTAGTAgggctagggcaacggttgc, for mutation of NF-Y motif at position (-38), GacggttggattttgaaggagAAGTAaggcgctcggagcggagagt. Expression plasmids for wild-type human p53 and mutants L22Q/W23S were gifts of C.Hurris (National Institutes of Health, Bethesda, USA) and were described in (Zhou et al., 1999). Expression plasmid for the p53 dominant-negative peptide (p53-DD) was a gift of M.Oren and was described in (Shaulian et al., 1992). Expression plasmid for p16 was kindly provided by R.Agami (Netherlands Cancer Institute). E2F-dTA expression plasmid pRcCMVE2F1-(1–363), encoding a dominant negative form of the E2F transcription factors, was as described in (Hofmann et al., 1996). dnNF-YA Expression plasmid NF-YA13m29, encoding a dominant negative form of the NF-Y transcription factor subunit A, was described in (Mantovani et al., 1994).

**Transfections and Reporter Assays**

HCT116 cells were plated at $3*10^4$ cells/well in a 24-well plate 48 hours before transfection. Cells were transfected with JetPEI transfection reagent (Polyplus Transfection), using 150 ng/well of luciferase reporter construct, 50 ng/well of pCMV-β-galactosidase expression vector for normalization of transfection efficiency, and appropriate expression plasmids for a total DNA amount of 1µg/well. The p53 expression plasmids were transfected at 10 ng/well. The p16, dnNF-YA and E2F-dTA expression plasmids were transfected at 300 ng/well. Cell extracts were prepared 48 hours after transfection, and luciferase and β-galactosidase activities were determined using commercial reagents and procedures (Promega). Statistical significance was determined by the paired t-test.

**RNA Preparation, cDNA Synthesis and qPCR**

Total RNA was extracted with Versagene RNA cell kit (Gentra Systems, Inc., Minneapolis, MN) and treated with DNase (Versagene DNase treatment kit, Gentra Systems, Inc.) following the manufacturer's instructions. A 2µg aliquot of the total RNA was reverse transcribed using MMLV-RT (Promega) and random hexamer primers (Roche Applied Science). Quantitative Real-time PCR (qPCR) was performed using SYBR Green PCR Master Mix (Applied Biosystems). The expression level for each sample was normalized to that of the GAPDH housekeeping gene in the same sample. Primer sequences were as follows: GAPDH, 5'-agcctcaagatcatcagcaatg-3' and 5'-cacgataccaaagttgtcatggat-3'; cdc20, 5'-gagggtggctgggttcctct-3' and 5'-cagatgcgaatgtgtcgatca-3'; CCNF, 5'-catctgcacccggtttatca-3' and 5'-cttccaaggcggagacga-3'; BIRC5, 5'-tcatccactgccccactga-3' and 5'-agaagaaacactgggccaagtc-3'; MAD2L1, 5'-gttggaagtttcttgttcatttgatct-3' and 5'-ggtcccgactcttcccattt-3'; CENPF, 5'-agaaagcagtcatgagtggtattca-3' and 5'-gcaggatatatgggctagtctttcc-3'; PRC1, 5'-acaaaccgaggaggaaatcttct-3' and 5'-caattcgtgccttcaactcttct-3'; Bub1b, 5'-tacactggaaatgaccctctggat-3' and 5'-tataatatcgttttctccttgtagtgctt-3'.

**Acknowledgments**

We thank all members of the Domany, Rotter and Pilpel labs for stimulating discussions. This research was supported by grants from the Israel Academy of Sciences and the Ben May Foundation (YP), the Leo and Julia Forchheimer Center for Molecular Genetics (YP), the FAMRI foundation (VR), the Ridgefield Foundation, and by the NIH (grant #5 POI CA 65930-06). VR holds the Norman and Helen Asher Professorial Chair in Cancer Research at the Weizmann Institute. ED is the incumbent of the Henry J. Leir Professional Chair. YP is an incumbent of the Aser Rothstein Career Development Chair in Genetic Diseases, and is a Fellow of the Hurwitz Foundation for Complexity Sciences.

**Figure Legends**

Figure 1 **A.** Outline of the malignant transformation process. Schematic representation of the spontaneous (young, senescent, immortal, tumorigenic, INK4A methylation) and induced (hTERT, H-Ras, p53 inactivation) modifications of the WI-38 cells along the process of malignant transformation. The stages chosen for microarray profiling are indicated by boxes with numerals corresponding to columns in the expression matrix shown in **B**. The time scale of the process is depicted by a horizontal axis, and the corresponding population doublings represented by PDLs. **B.** The normalized expression levels of the 168 genes in the proliferation cluster at twelve stages spanning the transformation process. Normalized expression level is color-coded according to the color bar on the right. The table below the matrix contains the following information on each sample: days in culture, geometric mean and standard deviation of expression level of the cluster's genes, doubling rate (cell cycle doublings/day) of cells at selected stages, activity of hTERT (designated as '+' for all samples following hTERT overexpression), activity of p53, as inferred from the application of its dominant-negative peptide, GSE56 ('-' indicates expression of GSE56). Here and throughout the paper the following cell line designations are introduced: cells are either young, or senescent; grow slow or fast; a sample name followed by 'G' denotes the application of GSE56; T before sample names indicates the presence of the immortalizing telomerase; R following the samples name indicates the insertion of Ras.

Figure 2 Motif positional bias in promoters of the proliferation cluster genes. The preferred window position of the five regulatory motifs, NFY.01, CDE, CHR, ELK1, and E2F.02 in the promoters of the proliferation cluster genes is shown. The CHR motif has also a clear strand bias (depicted by a directional arrow here), for more details see Fig. S6.

Figure 3 **A.** Graph depicting interaction between the five regulatory motifs measured by synergy and co-occurrence. A pair of motifs connected by a thick red line have a synergistic effect on expression and pairs connected by a thin blue line are significantly more highly co-occurring in individual

promoters. These interactions were computed using all genes on the array. In addition we calculated co-occurrence interaction given only the genes in the cluster (i.e. considering the total number of genes in the cluster and the number of cluster's genes containing each motif as a background), and found even there several significant interactions (blue dotted lines). Since for some of the motifs, the motif database contains multiple variants, we unified all variants of the same motif into one node, and an edge in the graph connects between two motifs if at least one of the variants of the two motifs are either synergistic or highly co-occurring. **B-C.** Combinogram analyses (Pilpel et al., 2001) using the five regulatory motifs during the transformation process (B) and HeLa cell cycle (C). All the varying genes in the respective arrays were used in the analysis. The upper and middle parts of the Combinogram are depicted as before (Pilpel et al., 2001), while the lower part is modified. Briefly, the middle section of the Combinogram depicts the motif composition of each gene set defined by Motif Combination (GMC, see Methods). Each vertical column represents a single GMC. A black square indicates that the particular motif is present in the promoters of all the genes in that GMC. A white square indicates that none of the genes in the GMC contain the particular motif. The top section of the graph shows the dendrogram analysis that assesses the similarity in expression profiles of each GMC using normalized Euclidean distances between the average expression profiles of the genes in the GMC as a measure of distance. The lower part of this modified Combinogram displays the mean expression profiles of the genes in each GMC, color-coded as in Fig. 1. The numbers of genes in each GMC appear below the dendrogram. The three main branch points in the dendrogram in **B** are represented as 1-3 (in circles). In the present analysis only genes that contain the motifs in their preferred distance relative to TSS (as depicted in Fig. 2) were considered, as genes that contain the motifs, yet away from the preferred location interestingly display no clear expression patterns (not shown). Since CHR motif has a strong strand bias in addition, and since genes that contain this motif yet on the non-preferred strand have a non-coherent expression profile (Fig. S6), for this motif we considered only genes that contain it in the preferred strand *and* distance from the TSS.

Figure 4 **A.** Normalized expression profiles of the tumor-suppressors p21 (brown line) and p16 (black line), along with a profile that represents the average of their profiles (green area), and a profile that represents the mean expression profiles of the genes in the proliferation cluster (blue area). **B.** Average expression profiles of all genes in the genome that contain Elk1 in their promoters (left) and the NFY and CHR motifs (right). Each bar represents an average over the samples that corresponds to low expression ($<-0.3$) denoted by 'L', and high expression ($> 0.2$) denoted by 'H' of the normalized mRNA expression levels of p21 and p16.

Figure 5 Three-way linkage between expression profiles, promoter architecture and tumor suppressor pathways. **A.** The entire expression matrix of the proliferation cluster genes, sorted with SPIN (Tsafrir et al., 2005), revealing an "elongated" shape for this cluster. B. Correlation coefficient between p21 and between p16 expression profiles to each of the genes in the cluster, color-coded according to color bar shown. **C.** Cumulative distribution of CHR, ELK1, and NFY along the sorted list of genes in **A. D.** Bars depicting main areas of density of regulatory motifs along the sorted expression matrix in **A.**, are based on cumulative appearance in **C.** and **E.** Arrows in the networks represent positive (green) or negative (red) interactions. Reviewed in: 1. (Ullrich and Schlessinger, 1990) 2. (Gille et al., 1995) 3. (Lin et al., 1998; Serrano. M et al., 1997; Zhu et al., 1998) 4,5,7,9-11 are reviewed in (Hahn and Weinberg, 2002; Sherr, 1996; Sherr and Roberts, 1999; Weinberg, 1995) 6. (el-Deiry et al., 1993) 8. (Yun et al., 2003); 12. Newly proposed interaction (this paper).

Figure 6 p53 represses proliferation cluster genes expression following DNA damage. Normal and GSE56-expressing WI-38 cells were treated with 0.2 µg/ml doxorubicin (dox) for 48 hours. mRNA levels for the indicated genes (y-axis) were measured by qPCR and normalized to the GAPDH housekeeping control.

Figure 7 **A.** The cdc20 promoter is repressed by wild-type p53 and by p16, but not by a transactivation deficient p53 mutant. Normalized luciferase activity of the *p-cdc20-luc* reporter in HCT-116 p53-/- cells. cdc20 promoter activity in the absence (control) or presence of either wild-type p53 (wt-p53), the L22Q/W23S p53 mutant (p53 22,23), or p16. **B**. cdc20 promoter is regulated by E2F and NF-Y. Normalized luciferase activity of *p-cdc20-luc* reporter in HCT-116 p53-/- cells. cdc20 promoter activity in the absence or presence of wild-type p53 and in the presence of either control vector (control), dominant-negative E2F1 (E2F-dTA), or dominant-negative NF-YA (dnNF-YA). Fold repression was calculated as the ratio of promoter activity in the absence of wt-p53 to the level in its presence, and was significantly lower in the presence of dominant-negative E2F1 and dominant-negative NF-YA compared to control (paired t-test p-values = 0.03 and 0.01, respectively). **C**. NF-Y motifs are important for p53-mediated repression of the cdc20 promoter. The wild type promoter construct *p-cdc20-luc* and a construct with both NF-Y motifs mutated (*mutant NF-Y 1+2*) were cotransfected into HCT116 p53+/+ cells in the presence or absence of a dominant-negative p53 (p53-DD). Fold repression was calculated as the ratio of promoter activity in the presence of p53-DD to control and was significantly lower for the *mutant NF-Y 1+2* construct (p-value = 0.005). Data represent the average of three independent experiments, each performed in triplicate.

**Supporting Information**

**Supplemental Data 1.** Microarray-derived expression levels of 168 genes of the proliferation cluster along the 12 time-points (each in duplicate) of the *in-vitro* transformation.

**Supplementary Figure 1.** Properties of the proliferation cluster.

**Supplementary Figure 2.** Expression profiles of 53 proliferation cluster genes during three human cell cycles (Whitfield et al., 2002).

**Supplementary Figure 3.** Evolutionary conservation of motifs and promoter architecture between human and mouse proliferation cluster-derived orthologous promoters.

**Supplementary Figure 4**. Negative correlation between the proliferation cluster genes expression level and the sum-gate of p16 and p21.

**Supplementary Figure 5**. An annotation-based procedure to determine a threshold score in assigning genes to a PSSM.

**Supplementary Figure 6**. Only genes that contain the CHR motif in its preferred strand and location display the proliferation signature.

**Supplementary Table 1.** Functional categories enriched in the proliferation cluster.

**Supplementary Table 2.** 5'-UTR analysis of the proliferation cluster genes.

**Supplementary Table 3.** Enriched motifs of the proliferation cluster are also over-represented in orthologous mouse promoters.

**Supplementary Table 4.** Promoter motif content of ~33,000 genes that are not represented on the Focus Array.

**Supplementary Table 5.** Over-representation p-values of all the MatInspector motifs in the proliferation cluster.

**Table I. Over-represented regulatory motifs in the proliferation cluster.**

| Motif | Number of genes containing motif among the proliferation cluster/entire array | Over representation p-value |
|---|---|---|
| NFY | 77/1574 | 9.92E-12 |
| E2F | 85/2617 | 3.57E-07 |
| CDE | 101/3073 | 2.20E-09 |
| ELK1 | 38/960 | 4.32E-05 |
| CHR | 10/63 | 5.43E-07 |
| CHR-NFY-CDE | 9/12 | 1.18E-13 |

The number of genes that contain the corresponding motif in their promoter among the proliferation cluster (out of 165 genes) and among the entire array (out of 8110 genes). A hyper-geometric p-value score was calculated in order to assess the extent to which a motif is over represented in significant location among the cluster's genes compared to the rest of the genes on the array.

**Table II. p53- and p21- dependent repression of the proliferation cluster genes expression.**

| | Fold Repression | | |
|---|---|---|---|
| Gene Symbol | HCT116 p53+/+ | HCT116 p53-/- | HCT116 p21-/- |
| Cdc20 | 2.6 | 1.2 | 0.7 |
| BIRC5 | 1.6 | 0.6 | 0.7 |
| NEK2 | 1.8 | 0.5 | 0.9 |
| Mad2L1 | 2.4 | 0.7 | 0.8 |
| Bub1B | 1.7 | 0.8 | 1.1 |
| PRC1 | 1.7 | 0.9 | 0.9 |
| CENPF | 1.7 | 0.7 | 1.2 |

HCT-116 cells containing wild type p53 and their p53-null and p21-null derivatives were treated with 0.2 µg/ml doxorubicin for 48 hours. The expression level for each gene was measured by qPCR and normalized to GAPDH expression level. The numbers in the table represent fold repression, calculated for each gene as the ratio of expression level in non-treated cells to doxorubicin-treated cells.

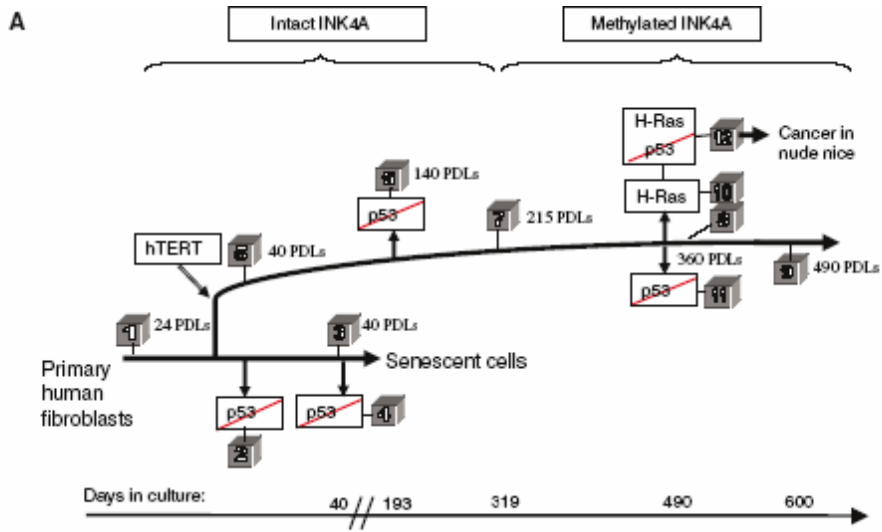

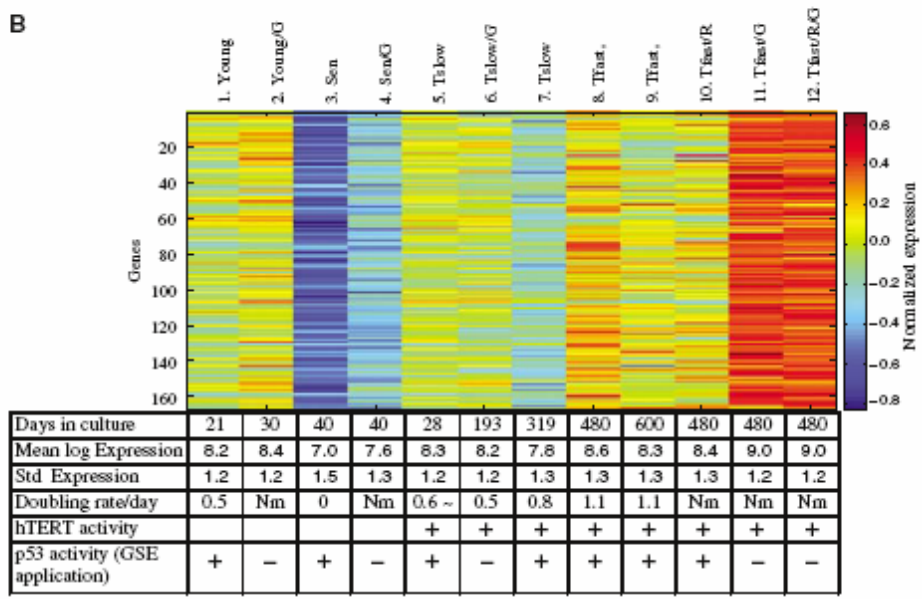

Fig 1

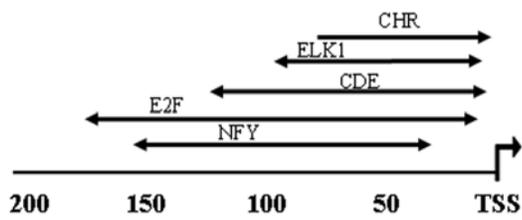

Fig 2

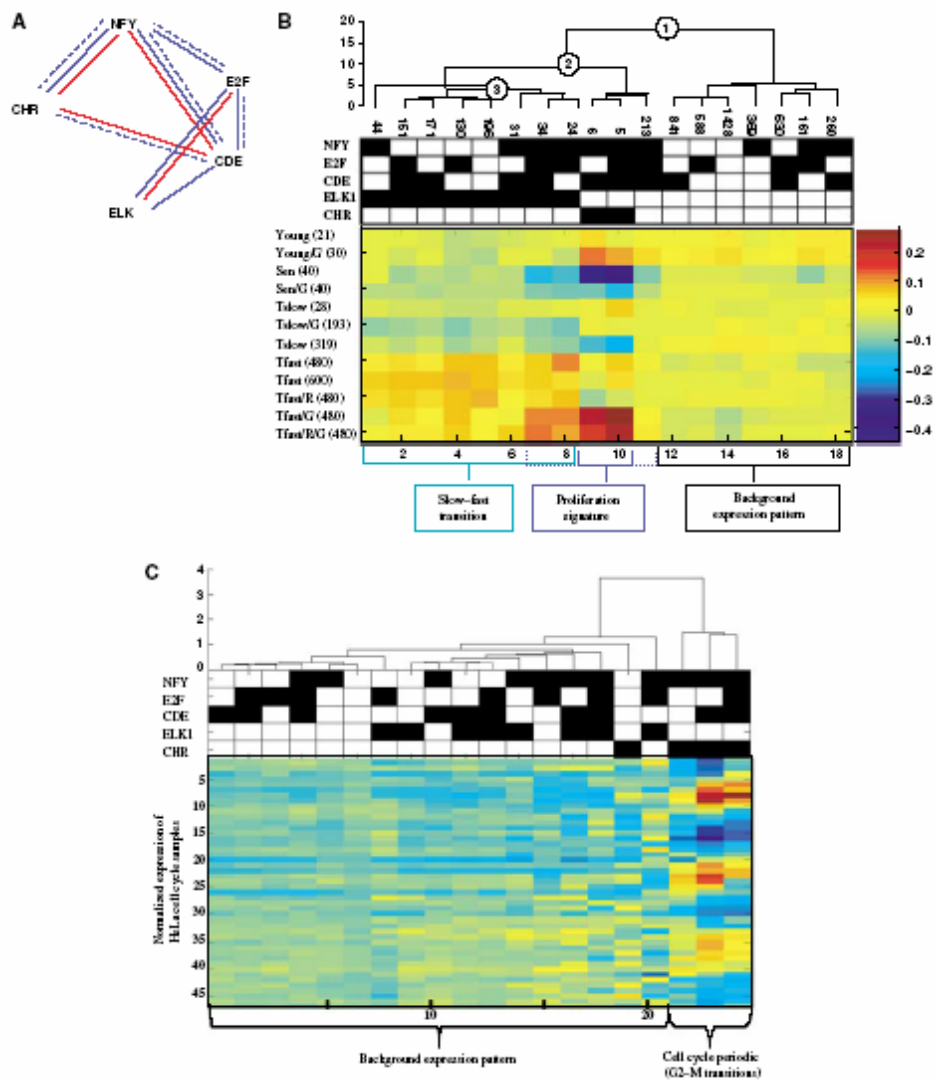

Fig 3

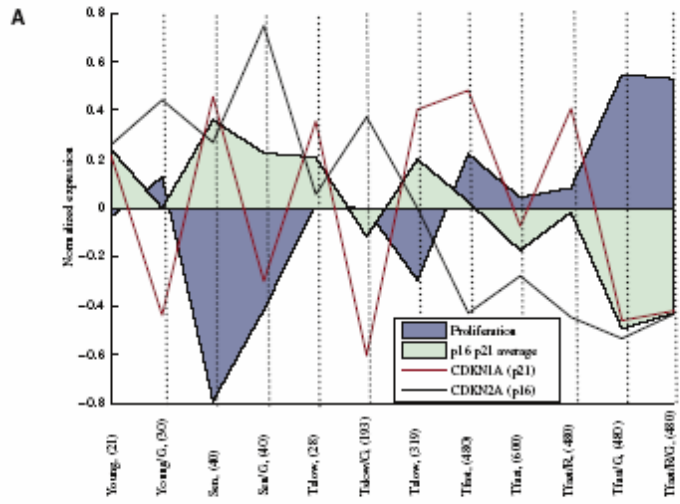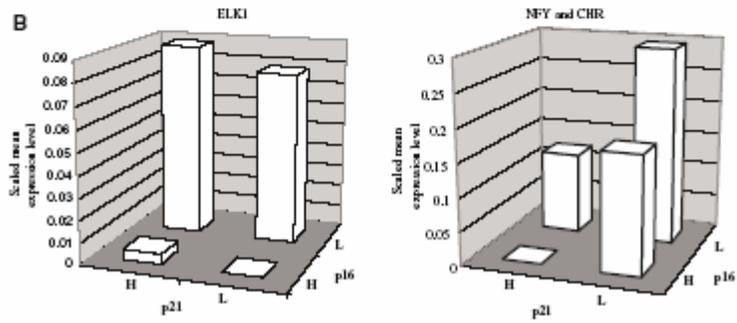

Fig 4

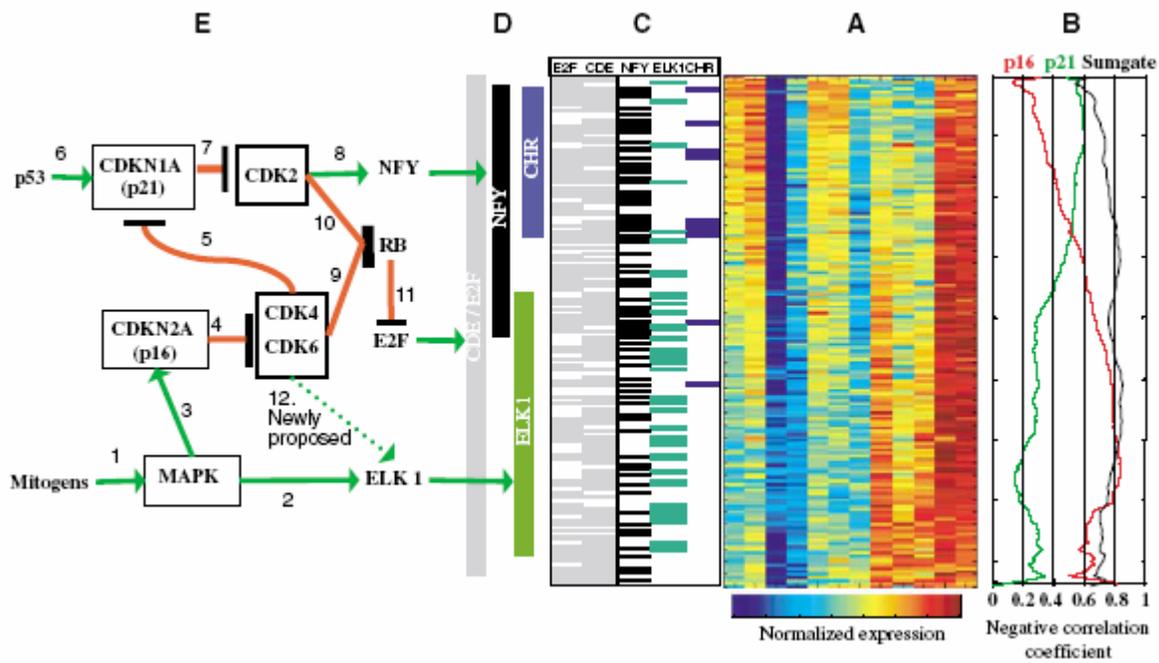

Fig 5

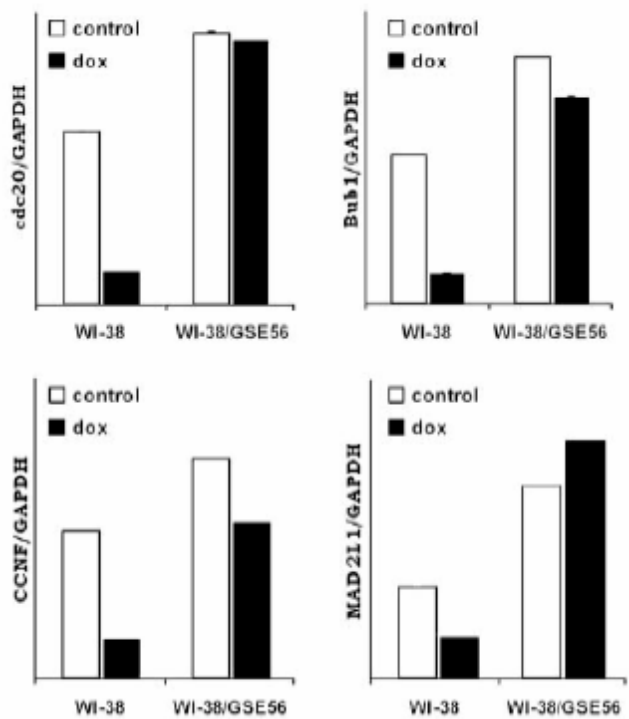

Fig 6

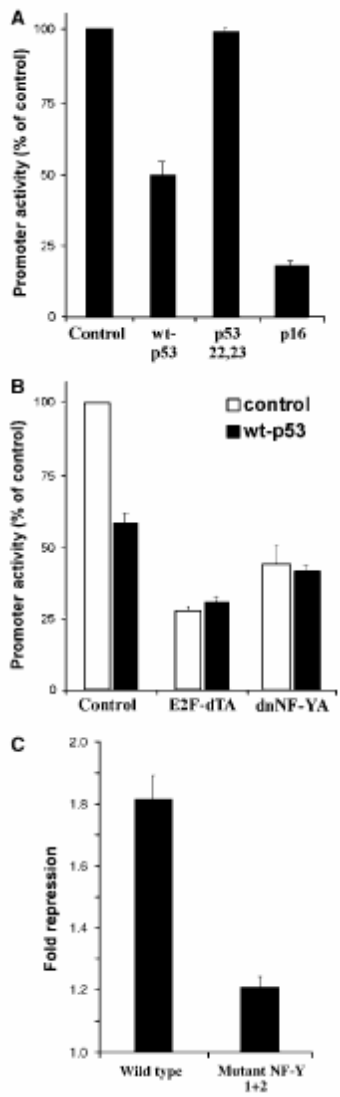

Fig 7